# 歴史に刻まれた巨大宇宙天気現象

早川　尚志（大阪大学/ラザフォード・アップルトン研究所）

海老　原祐輔（京都大学）

## 1 序論

　太陽面での爆発の内でも極端に大きなものがしばしば激甚宇宙天気災害を引き起こし、電力網や通信網、人工衛星といった各種近代インフラに甚大な影響を及ぼすことは既に第1章や第2章3節で論じられた通りである。とりわけ、観測史上最大とされるキャリントン・イベント（Tsurutani *et al.* 2003; Cliver and Dietrich 2013）に相当する極端宇宙天気現象が今日発生した場合、現代社会の依拠する各種インフラへの影響は甚大なものになると評価されている（Baker *et al.* 2008）。ロンドンのロイズ保険組合の報告書はこのような宇宙天気現象の脅威について「長きにわたる電力供給の喪失は社会を19世紀の慣習へ逆戻りさせかねない」と警鐘を鳴らしている（Hapgood 2010）。

　幸いこのような極端宇宙天気現象の発生頻度は低く、統計的な検討から100年に一度程度と評価する研究が多い（Riley *et al.* 2018）。一方でこのような低い発生頻度は極端宇宙天気現象を定量的に検討する上で大きな障壁となっている。太陽面上の爆発現象やその結果生じる磁気嵐について体系的なデータベースが整備されているのは直近の半世紀強に過ぎないからである。そのため、キャリントン・イベントのような低頻度の極端宇宙天気現象を理解するためには、極端宇宙天気現象の代替データによる長期的な調査を行う必要がある。

　過去の宇宙天気現象の研究にあたって注目すべきは太陽爆発の発生源となる太陽黒点の観測記録、その結果生じた磁気嵐を記録した地磁気観測記録、そして磁気嵐の際に生じるオーロラの観測記録である。磁気嵐の規模はDst指数の最大振幅によって直接的に評価されることが多い。Dst指数は中緯度の4観測所で測定された地磁気変動の平均で（*e.g.*, Sugiura and Kamei 1991）、近代的な地磁気観測が無い期間についてはDst指数を求めることはできない。オーロラ帯は磁気嵐がおこると低緯度に移動し（Akasofu 1964）、オーロラ帯の低緯度境界の磁気緯度はDst指数とよい相関があることが経験的に知られている（Yokoyama *et al.* 1998）。この関係を利用すると、オーロラ帯の低緯度境界から磁気嵐の規模が推定することができる。

　本節ではオーロラの記録や黒点スケッチに基づいて復元した過去の宇宙天気現象についての最近の成果を紹介する。

## 2 キャリントン・イベント

　観測史上最大級の宇宙天気現象は1859年9月のキャリントン・イベントと考えられる。9月1日の観測史上初の白色光フレア（Carrington, 1859; Hodgson, 1859）から遅れること17.6時間後、大規模な磁気擾乱と低緯度オーロラが発生した。この際、ボンベイで−1600 nT に及ぶ地磁気水平成分の擾乱が観測され（Tsurutani *et al.*, 2003）、ホノルルに至るまでオーロラが観測されたことから（Chapman, 1957）、しばしばこの宇宙天気現象は観測史上最大の宇宙天気現象として考えられる（*e.g.*, Tsurutani *et al.* 2003; Cliver and Dietrich 2013）。

　一方、キャリントン・イベントが発生したのは1859年と現代機器観測の始まりよりも相当前のことである。それ故、当該イベントの規模推定にはばらつきがある点は注意が必要である。キャリントン・イベントのフレアのX線強度は、太陽面爆発時のX線到達による電離層の電気伝導度変化によって生じるSFE（solar flare effect）の振幅によって推定されてきた。イギリスのグリニッジやキューで計測されたSFEに伴う水平成分の振幅が ≈ 110 nTに及ぶことから（図1）、キャリントン・フレアのX線強度は2003年11月に発生した巨大フレアとの比較で当初 > X10 と推定され（Cliver and Svalgaard 2004）、その後さらなるフレアX線強度とSFEの統計的比較検討によって ≈ X45と推定されている（*e.g.*, Boteler 2006; Cliver and Dietrich 2013; Curto *et al.* 2016）。

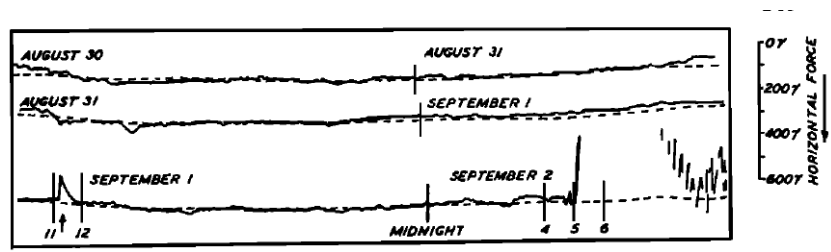

図**1**．キャリントン・イベント時のキュー観測所のSFEと磁気嵐（Bartels, 1937）

　磁気嵐は中低緯度の地磁気水平成分が数日間にわたり減少することで認識され、その規模は中低緯度で測定した地磁気水平成分の振幅により評価される。現在、キャリントン・イベント時に中低緯度で測定された地磁気水平成分のデータとしてはボンベイで得られたものが唯一のものとされている。数日間にわたる地磁気の減少は主に地球を取り囲むように流れる赤道環電流が発達することでおこる。赤道環電流という名から想像される構造とは異なり、その分布は軸対称ではなく、夜側で強く昼側で弱いという地方時依存性を示している（Ebihara *et al.*, 2002; Le *et al.*, 2004）。ボンベイでの測定によると、キャリントン・イベント時における地磁気水平成分の変動は最大で−1600 nTにも及ぶ（Tsurutani *et al.*, 2003）。−1600 nTもの地磁気変動が中低緯度で測定されることは極めて珍しく、その原因をめぐっては赤道環電流であるという考え

（Tsurutani *et al*., 2003）と電離圏電流によるという考え（Kamide and Akasofu）があり対立している。もしボンベイで記録された地磁気の減少の原因が赤道環電流であるならば、ボンベイが昼側に位置していたことと赤道環電流の地方時依存性を考慮すると最小Dst値は−1760 nTになる可能性が指摘されている（Tsurutani *et al*., 2003）。ただしDst値は1時間値であるという定義に従うと、最小Dst値は−900 (+50, −150) nTになるとの推定もある（*e.g.*, Siscoe *et al*. 2006; Gonzalez *et al*. 2011; Cliver and Dietrich 2013）。同時代の地磁気観測データは幾つかあるが、その多くは振り切れていることから、当時の電流系の全容を明らかにすることは容易でない。ボンベイの地磁気変化に局所的な電流系の影響が混在している可能性があり、今後も実際の観測データに基づく議論が必要である（*e.g.*, Love *et al*. 2019a, 2019b）。

　磁気嵐がおこるとオーロラオーバルが拡大することが知られ（Akasofu 1964）、その低緯度境界の磁気緯度とDst指数の間にはよい対応関係があることが知られている（Yokoyama *et al*. 1998）。キャリントン・イベント時の低緯度オーロラの低緯度境界はLoomis（1860）などによる同時代の研究以来、盛んに論じられてきた。オーロラの可視範囲はホノルルやサンティアゴの記録に基づいて磁気緯度（MLAT）にして23°（Tsurutani *et al*. 2003）、或いはカリブ海の航海記録に基づいて当該イベント時のオーロラ可視範囲が18°まで及んだとされている（Green and Boardsen 2006）。その後検討された東半球やメキシコの一連の記録も合わせて、キャリントン・イベント時のオーロラ可視範囲は図2の様になる（Hayakawa *et al*. 2019e）。この記録からオーロラオーバルの低緯度境界を復元するためにはオーロラ観測地点の磁気緯度だけでなく仰角も検討する必要がある（図3）。地球磁場を双極子磁場で近似し、赤色オーロラの上端高度を400 kmと仮定すると（Roach *et al*. 1960; Ebihara *et al*. 2017）、仰角情報が含まれるホノルルの記録からオーロラオーバルの低緯度境界の不変磁気緯度（invariant latitude, 以下ILAT）は28.5°/30.8°と見積もることができる（Hayakawa *et al*. 2018e, 2019e）。ただし、ホノルルの記録には日付の不定性があり、注意を要する。不変磁気緯度は磁力線に沿って一定で、双極子磁場のもとでは磁力線と地表が接する地点に磁気緯度と合致する。荷電粒子は磁力線に沿って動きやすいという性質があることから、地球磁場に捕捉された荷電粒子やその結果として生じるオーロラの分布を調べる上で有用である。

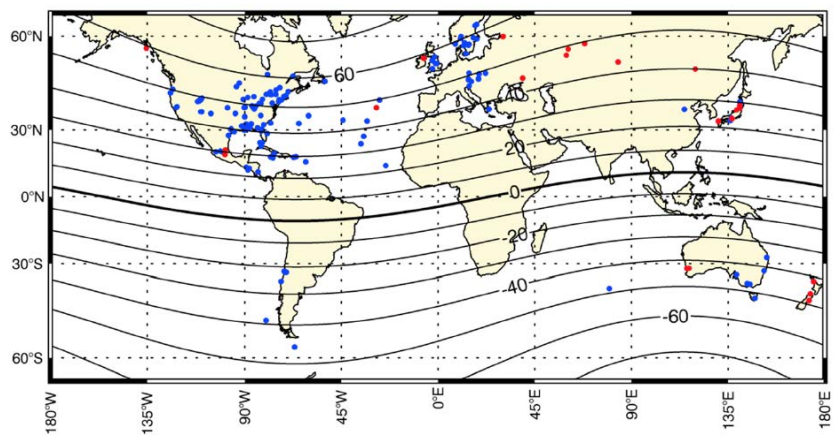

**図2．**キャリントン・イベント時（1859年9月1/2 – 2/3日）周辺のオーロラ可視範囲（Hayakawa *et al.*, 2019e）。青点は既知の観測地点、赤点はHayakawa et al. (2019e) を示す。黒のコンターは磁気緯度を表す。

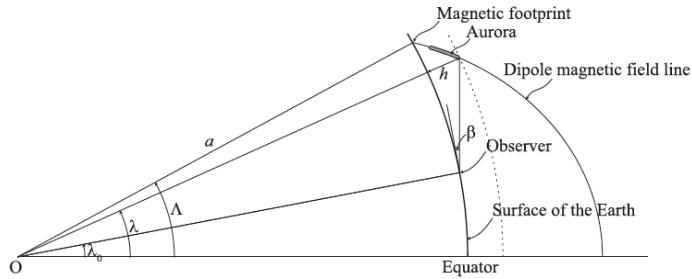

**図3．**オーロラ可視範囲とオーロラオーバルの低緯度境界（Hayakawa *et al*., 2018e）$\lambda_0, \lambda, \Lambda, a, h, \beta$はそれぞれ観測者の磁気緯度、オーロラ上端の磁気緯度、不変磁気緯度(ILAT)、地球半径、オーロラ上端の高さ、観測者から見たオーロラ上端の仰角である。

　このようにキャリントン・イベントは少なくともフレアのX線強度（≈ X45）、磁気嵐規模（Dst ≈ −900 (+50, −150) nT）、オーロラオーバルの低緯度境界（≈ 30.8° ILAT）と定量的に復元され、史上最大規模の宇宙天気現象と位置付けられている（Tsurutani *et al*. 2003; Cliver and Dietrich 2013; Hayakawa *et al*. 2018e）。この他、キャリントン・イベントはグリーンランド氷床コアの窒素酸化物濃度から史上最大級のSEPイベントと考えられたこともあったが（McCracken *et al*. 2001; Smart *et al*. 2006）、同年代の$^{10}$Be同位体比や他の氷床コアの窒素酸化物濃度との比較検討の結果、現在ではこの推定には疑義が呈されている（Usoskin and Kovaltsov 2012; Wolff *et al*. 2012; Schrijver *et al*. 2012）。

　この当時の太陽面の様子は特にCarrington（1859）の部分的な黒点スケッチから知られているが、Carrington本人を含めた同時代の複数の観測者の黒点スケッチの多くは各地の文書館に未公刊史料として所蔵されている（図4）。これらのスケッチの分析から、キャリントン・イベントを起こした黒点が8月28日は太陽の東のリムに現れ、8月31日から9月1日にかけて太陽面の中央を通過し、9月7日頃に太陽の西のリムを越していったことが明らかになる。8月の宇宙天気現象時の黒点の位置は必ずしも地球向けのCMEを起こしやすい場所にはなかったが、9月1日のフレア時点ではちょうど地球の正面に巨大黒点が来ていたため、CMEが地球を直撃しやすい位置にあった点は特筆に値する（Hayakawa *et al*. 2019e）。

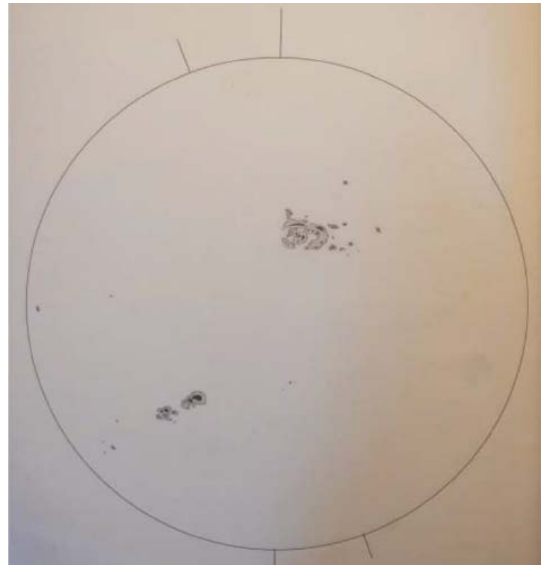

**図4．**キャリントン・イベント時のキャリントン直筆の黒点スケッチ（英国王立天文学会蔵）。リムとコントラストを強調し、投影法の方向を補正してある（Hayakawa *et al*. 2019e）。

# 3 近代観測に捉らえられた極端宇宙天気現象

　一方でキャリントン・イベントが近代観測中にどの程度特異な宇宙天気現象だったかについては、依然慎重な議論が必要になる。既に近年、いくつかキャリントン・イベントの「なり損ない」が報告されているためである。2012年7月に巨大なCMEが地球とは反対方向に放出されたことは記憶に新しいが、このCMEが地球に衝突した場合、地軸及び地磁気極の向きによってはDst ≈ −1182 nTに及ぶ極端磁気嵐が発生した可能性が指摘される（Baker *et al.*, 2013; Ngwira *et al.*, 2013）。また、1972年8月の史上最速のCMEは地球磁気圏に14.6時間で到達したが（Russel *et al.*, 2013）、この時はCMEの磁場が北向き気味だったことなどもあって磁気嵐規模はDst ≈ −125 nTと慎ましやかなものに留まった（Tsurutani *et al.* 2003; Knipp *et al.* 2018）。

　近代観測に基づき、キャリントン・イベントはどの程度特異だったのか考えてみたい。幸い地磁気変動やオーロラの体系的な近代観測は1830年代には始まっており（*e.g.*, Chapman and Bartels 1940）、2世紀弱の時間幅で検討を行うことが可能である。興味深いことにChapman（1957）は1859年9月のキャリントン・イベントの他に、1872年2月にボンベイで、1909年9月にシンガポールで、1921年5月にサモアでオーロラが見えたことを指摘し、これら三つの宇宙天気現象をキャリントン・イベント同様に "outstanding aurora" と位置付けている。これらの観測地点はいずれも磁気緯度が20°を下回ることから、オーロラ可視範囲やオーロラオーバルの赤道側境界、各々の磁気嵐の規模について議論が行われた（*e.g.*, Hayakawa *et al.* 2019e）。

　1872年2月4日の磁気嵐は太陽サイクル11の極大からの下降期に発生した。オーロラはボンベイ以外では北インド、西ヨーロッパ、インド洋、中東、南北アフリカ、東アジアなど磁気緯度 ≈ 19° に至るまでの各地で観測された（Silverman 2008; Hayakawa *et al.* 2018d）。ボンベイの磁気緯度は10°程度で、この中でも際立って低緯度側に位置する一方、オーロラ記録の記述は精密で、Silverman（2008）は磁気嵐に関係したオーロラと独立してsporadic auroraがボンベイで観測された可能性を指摘した。一方で、東アジアのオーロラ記録の分析の結果、オーロラが上海（磁気緯度 ≈ 20°）で天頂まで広がっていたことが判明し、インド側でも同程度にオーロラオーバルが広がっていたと仮定すると、ボンベイでも地平線から仰角にして10–15°程度に至るまでオーロラが広がっていた可能性が指摘される（Hayakawa *et al.* 2018d）。このオーロラ観測はボンベイで磁気擾乱がピークに達していた時期と合致し、観測記録から磁気擾乱は＜−830 nT程度に至っていたと考えられる。

　1909年9月の25日の宇宙天気現象は太陽サイクル14の下降期に発生した。この宇宙天気イベントにまつわるフレアは11–12.5時（UT）に発生し、白色光フレアやSFEとして英国各地の観測者に記録された（Lockyer 1909; Hayakawa *et al.* 2019a; Love *et al.* 2019a）。このフレアに伴うCME

は、その24.75時間後に地磁気擾乱を引き起こした。フレアの伝播時間やSFEの振幅からこのフレアの規模は ≥ X10と推定されている（Hayakawa *et al.* 2019a）。また、この磁気嵐の規模はモーリシャス、サンフェルナンド、ヴィエケス、アピアの磁力観測に基づいてDst ≈ −595 nT と復元されている（Love *et al.* 2019a）。この磁気嵐の主相から回復相にかけて夜側に位置していた日本やオーストラリアでオーロラが広く観測され、北海道でオーロラが天頂に見えた他、松山でオーロラが仰角30°までオーロラが観測されたことから、オーロラオーバルの低緯度境界は32° ILAT程度と復元されている（Hayakawa *et al.* 2019a）。

　1921年5月14/15日の宇宙天気現象は太陽サイクル15の下降期終盤、太陽赤道近くに現れた巨大黒点によって引き起こされた（Newton 1948）。この磁気嵐にまつわるフレアのタイミングには諸説あるが（Lefèvre *et al.* 2016; Love *et al.*, 2019b）、5月12–20日の間に六つの地磁気嵐急始（SSC, sudden storm commencement）が記録されていることから（Newton 1948）、この当時複数のフレアに際して放出された一連のCMEが惑星間空間を一掃し、巨大磁気嵐の発生しやすい状況を作っていた可能性がある（Love *et al.* 2019b）。この内、5月14/15日に生じた磁気嵐はとりわけ大規模で、南太平洋のサモア島に至るまでオーロラが見え（Silverman and Cliver 2001）、ニューヨークなどでの電話線障害や鉄道駅での火事などとの関係が議論されている（Hapgood 2019; Love *et al.* 2019b）。この磁気嵐の規模はアピア、ワゼルー、サンフェルナンド、ヴァスラスの磁力観測に基づいてDst ≈ −907 ± 132 nT（Love *et al.*, 2019b）、オーロラオーバルの低緯度境界はアピアの観測記録に基づいて27.1° ILAT（Hayakawa *et al.*, 2019f）と推定されている。

　このような極端宇宙天気現象と1989年3月の極端宇宙天気現象の磁気嵐規模とオーロラ低緯度境界を表1にまとめた。キャリントン・イベントは少なくとも推定したDstとオーロラオーバルの低緯度境界の点では、あくまで1872年2月や1921年5月の極端宇宙天気現象に相当する程度のもので、観測史上最大の巨大イベントではあったとは言いがたい。

**表1．** 近代観測中の宇宙天気現象のオーロラ可視範囲及びオーロラオーバルの低緯度境界と磁気嵐の規模推定値（Hayakawa *et al.* 2019f）。磁気嵐の規模推定値の内、*が付されているものは、低緯度の4観測所の平均値ではなく、ボンベイの磁力観測に基づく推定値による。

| 日付 | | | オーロラ可視範囲低緯度境界（MLAT） | オーロラオーバル低緯度境界（ILAT） | 磁気嵐規模（Dst: nT） |
| 年 | 月 | 日 | | | |
|---|---|---|---|---|---|
| 1859 | 8 | 28/29 | 20.2 | 36.5 | ≥ −484* |
| 1859 | 9 | 1/2 | 20.5/21.8 | 28.5 / 30.8 | ≈ −850-−1050* |
| 1872 | 2 | 4 | 10.0 / 18.7 | 24.2 | < −830* |
| 1909 | 9 | 25 | 10.0 / 23.1 | 31.6 | −595 |
| 1921 | 5 | 14/15 | 16.2 | 27.1 | −907 ± 132 |
| 1989 | 3 | 13/14 | 29 | 35 / 40.1 | −589 |

## 4. 近代観測と歴史文献の狭間に記録された極端宇宙天気現象

　近代観測（1830年代〜）の時間幅を越えて過去の宇宙天気現象を検討する場合、手がかりになるのは年輪や氷床コア中の放射性同位体比の変動や歴史文献中のオーロラ・黒点記録である（*e.g.*, Vaquero and Vázquez 2009; Usoskin 2017）。この内前者はSEPの同定（*e.g.*, Beer *et al.* 2012; Usoskin and Kovaltsov 2012; Barnard *et al.* 2018）に用立てられるが、歴史文献もオーロラ記録によるCMEの同定（*e.g.*, Willis and Stephenson 2000; Hayakawa *et al.* 2017a, 2017b）、黒点記録による出発点としての巨大黒点の同定などの面で重要になる（*e.g.*, Willis and Stephenson 2001; Willis *et al.* 2005; Hayakawa *et al.* 2017e）。

　このような宇宙天気現象の内、特に太陽黒点の近代観測の始まった1610年以降に記録されたものは、歴史文献と近代観測を照合しつつ検討を行うことが可能である。例えば1859年9月や1872年2月に発生した宇宙天気現象の際には東アジアでもオーロラが観測され、「赤氣」「白氣」「赤光」などの形で記録されている（Willis *et al.* 2007; Hayakawa *et al.* 2016b, 2018d）。また1769年11月や1917年2月に出現した巨大黒点は「黒子」「黑氣」などとして記録されている（Willis *et al.* 1996a, 2018; Hayakawa *et al.* 2019c）。

　以上のような用語を手掛かりに歴史文献を調べると、地磁気変動の近代観測が始まる以前の極端宇宙天気現象がいくつか浮上してくる。その内代表的なものは1770年１月、9月及び1730年2月に記録された宇宙天気現象であろう。1770年1月にはイベリア半島やカナリア諸島に及ぶまでオーロラが観測され、同時期に黒点数のピークの見られる点が指摘されている（Schröder 2010; Carrasco *et al.* 2018）。

　同年9月には東アジア一帯でオーロラが9晩近く連続して記録され、その可視範囲は南北半球で各々洞庭湖（18.9° MLAT）やティモール沖（−20.5° MLAT）に及んでいる（図5; Willis *et al.* 1996b; Hayakawa *et al.* 2017e）。また、9月17日のオーロラは月夜のように明るいとも記述されたが、この原因は低エネルギーの電子が大量に降下した結果だと考えられる（Ebihara *et al.* 2017）。9月17日のオーロラの低緯度境界の復元については依然議論があり（Ebihara *et al.* 2017; Kataoka and Iwahashi 2017）、今後当時の星座の位置や同時代に基づいて更に慎重な議論を行う必要がある。当時の太陽表面に目を転じると、独ニュルンベルクのシュタウダッハ（J. C. Staudach; Arlt 2008）の黒点スケッチに巨大黒点が記録されており、この巨大黒点が一連の極端宇宙天気現象を引き起こしたと考えられる（Hayakawa *et al.* 2017e; 図5）。

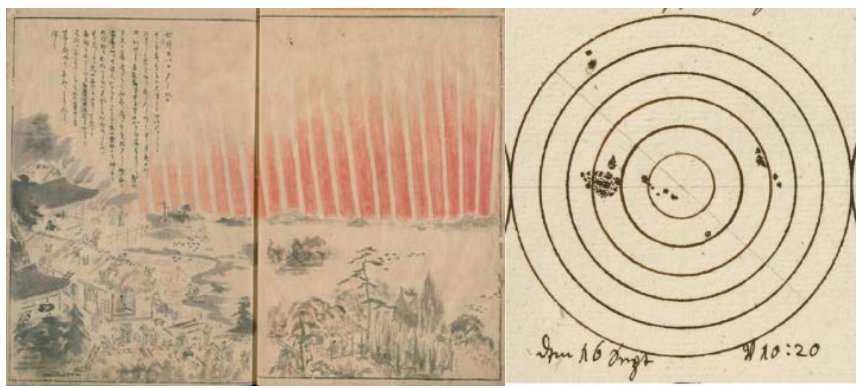

**図5．** 1770年9月17日に名古屋で観測されたオーロラ（国立国会図書館所蔵）と同9月16日の黒点スケッチ（ポツダム天体物理天文台所蔵）

　1730年2月にも東アジア一帯でオーロラが観測され、京都（25.4° MLAT）までオーロラが観測されたことが記録されている。同時代の黒点スケッチの検討の結果、1730年前半に黒点数が大きく盛り上がっていることが確認されており、当時太陽活動が相当に活発になっていたことが窺える（Hayakawa *et al*. 2018c）。

## 5 歴史文献で遡る宇宙天気現象の歴史

　このように、過去の宇宙天気現象は歴史文献にオーロラ記録として記録されることがあり、磁気嵐の規模推定の手がかりになる。このような歴史文献はとりわけ、より低頻度で大規模な宇宙天気現象の検討を行う上で重要である。例えば、近年の年輪や氷床コアの放射性同位体比の分析から、前660年、774/775年、992/993 – 993/994年に観測史上最大級の宇宙線イベントが確認されており、観測史上類を見ない極端なSEPを伴った大規模な宇宙天気現象が発生したことが指摘されている（*e.g.,* Miyake *et al*. 2012, 2013; Usoskin *et al*. 2013; Mekhaldi *et al*. 2015; Büntgen *et al*. 2018; Uusitalo *et al*. 2018; 3.4.1章参照）。一方でこれらの極端宇宙天気現象はいずれも地磁気（1830年代〜）や太陽黒点（1610年〜）の近代観測の始まり（Chapman and Bartels 1940; Owens 2013; Clette *et al*. 2014）の遥か以前に発生しており、周囲の宇宙天気現象の調査には近代観測以前の歴史文献を当たる必要がある。

　歴史文献に記録されたオーロラ記録は、前6世紀まで日付単位で遡ることが可能である。現状日付単位で記録されたものとしてはバビロン天文日誌に記録された前567年3月12/13日のオーロラ記録が現存最古であると考えられる（Stephenson *et al*. 2004; Hayakawa *et al*. 2016c; 図6a）。これに対し、聖書のエゼキエルの幻視を最古のオーロラ記録とする説もあるが（*e.g.,* Siscoe *et al*. 2002; Silverman 2006）、依然慎重な文献学的精査が必要である。解釈の当否はさておき、この幻視の日付は記述を信用するなら、前594年7月12/13日もしくは前593年7月30/31日に比定され

る（Hayakawa *et al.* 2019f）。

なお、既に前7世紀からアッシリアでは占星術のための天文観測が行われており（Hunger 1992）、その中には少なくとも3点オーロラを思わせる記述が見受けられる（図6b）。これらの占星術観測で日付が厳密に記録されているものは少ないが、当該の記録群については観測した学者の活躍年代から各々、前679 – 655年、前677 – 666年、前679 – 670年と観測年代を絞り込むことが可能である（Hayakawa *et al.* 2019f）。

一方オーロラの図像史料は現状771/772年の『ズークニーン年代記』のヴァティカン写本（MS Vat.Sir.162; 図6c）の図像が現存最古と考えられる。一般に図像の模写は黒点スケッチの例にも見える通り文字記録の模写以上に困難なため（Hayakawa *et al.* 2018a; Fujiyama *et al.* 2019）、図像が自筆か複写か、はたまた伝聞によるものかについては慎重な検討を要する（Hayakawa *et al.* 2017b）。

a)

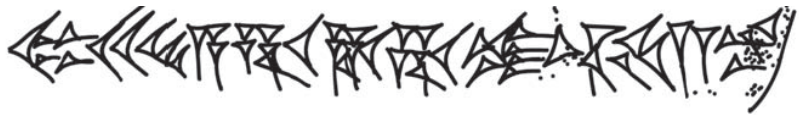

b)

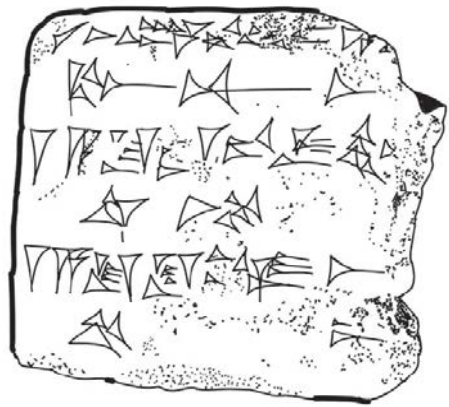

c)

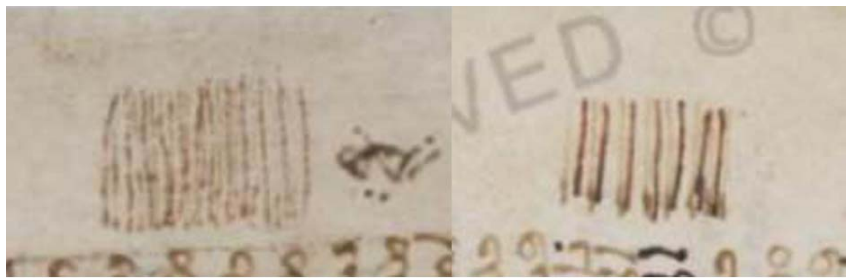

**図6．** 現存世界最古のオーロラ記録と図像史料：（a）前567年の『バビロン天文日誌』の模写、（b）前679 – 655年のアッシリア占星術レポートの模写（Hayakawa *et al.* 2019f）、（c）771/772年、773年6月の『ズークニーン年代記』のオーロラ図像（MS Vat.Sir.162, ヴァティカン教皇庁図書館（Biblioteca Apostolica Vaticana）所蔵; 三津間・早川 2017）

　オーロラと同様に黒点についても肉眼で観測された巨大黒点の記録が歴史文献中に散見される。現状、肉眼黒点の観測記録は少なくとも前165年の漢の記録まで遡ると考えられる（Yau and Stephenson 1988; Vaquero and Vázquez 2009; *c.f.*, Bicknell 1967）。前近代の肉眼黒点観測の大多数は東アジアで記録されているが、少数ながら東アジア以外での記録が残っていることもある（Yau and Stephenson 1988; Vaquero and Vázquez 2009; Hayakawa *et al.* 2017c）。実際に現存最古の黒点図像はイングランドの『ウースター年代記』（MS Oxford CCC 157）に記録された1128年12月8日のもので（図7）、『高麗史』に記録された5日後の赤氣の記録との比較から、大規模な宇宙天気現象を起こしたと考えられている（Willis and Stephenson 2001）。

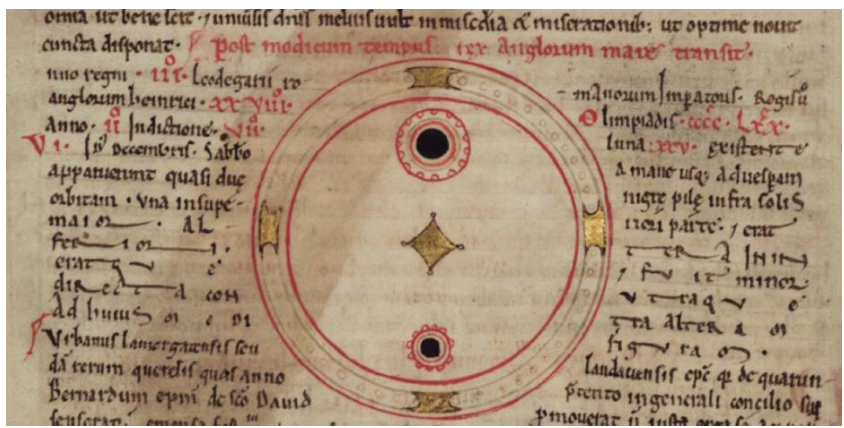

**図7.** 現存史上最古の黒点画像（MS 157, オックスフォード・コーパス・クリスティ・カレッジ所蔵）

## 6 歴史文献に記録された宇宙天気現象

　このように、歴史文献の黒点記録やオーロラ記録を精査すると、宇宙天気現象を過去20世紀以上に亘って復元することが可能である。例えば、本邦で記録された現存最古の天文記録と考えられるのは日本書紀に記録される620年の「赤気」であり（谷川・相馬 2008）、一般にこれはオーロラとして解釈されている（*e.g.,* 神田 1933）。奇しくも今年はいわば本邦の宇宙天気観測の1400周年の節目に当たる。その形は「雉尾」に例えられており（*e.g.,* 神田 1933）、雉の尾羽が開いているか否か次第で、水平方向に広がるアーク状のオーロラもしくは扇状に広がったレイ構造を思わせる（図8）。この際注意すべきは同時代の磁極の位置である。当時の地磁気極はユーラシア寄りに位置していたと考えられ、CALS3k4bモデル（Korte and Constable 2011）によると飛鳥の磁気緯度は35° MLATである。現在の北海道と同程度の磁気緯度上で、現在の同地点（飛鳥）と比べ、よりオーロラが見えやすい位置にあった（Hayakawa *et al.* 2017d）。しかし、北海道におけるオーロラ光学観測でアーク状のオーロラが記録されたことはないのに対し（Shiokawa *et al.* 2005）、扇状に開いたオーロラが1770年に京都など（~ 25° MLAT）で確認さ

11れた事例もあることから（Ebihara *et al*. 2017; Kataoka and Iwahashi 2017）、この日本書紀が描くオーロラの形状は雉が尾羽を開いたときに見えるような扇状のオーロラであったと解釈することが現状より妥当に思われる。

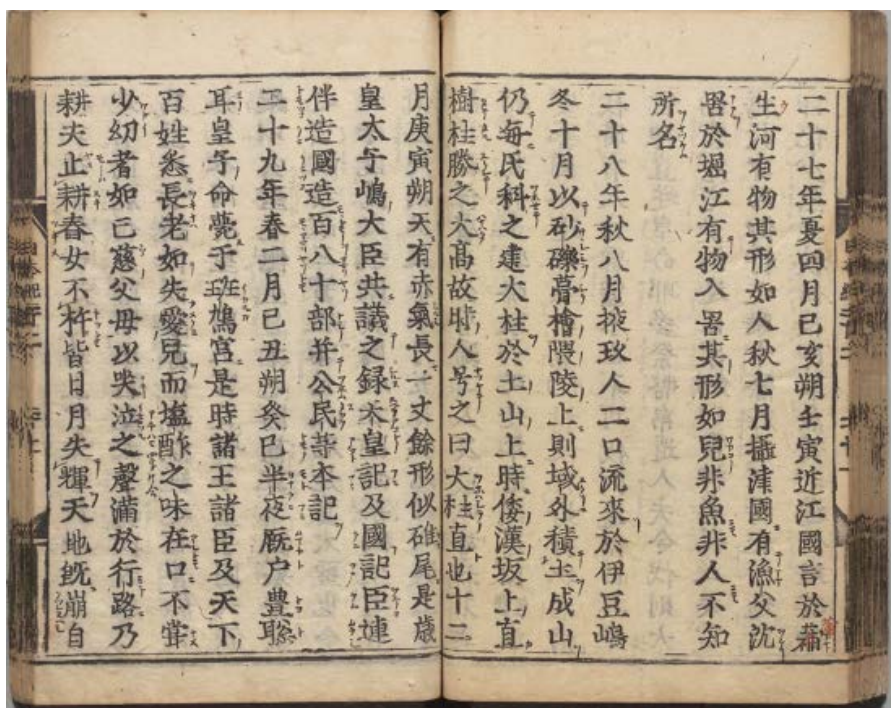

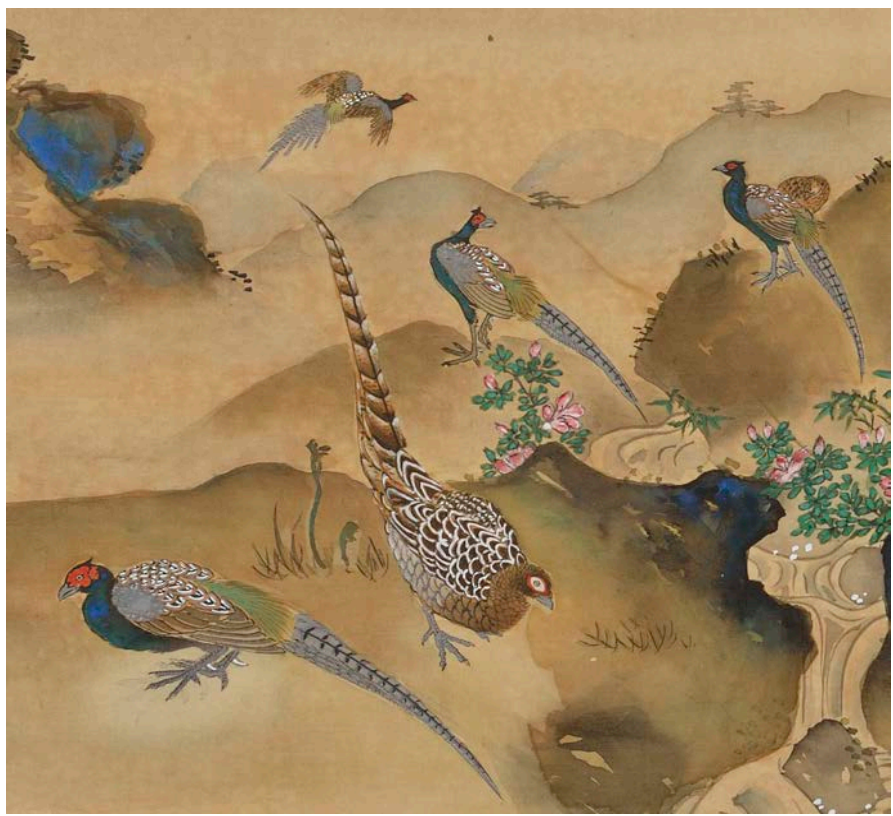

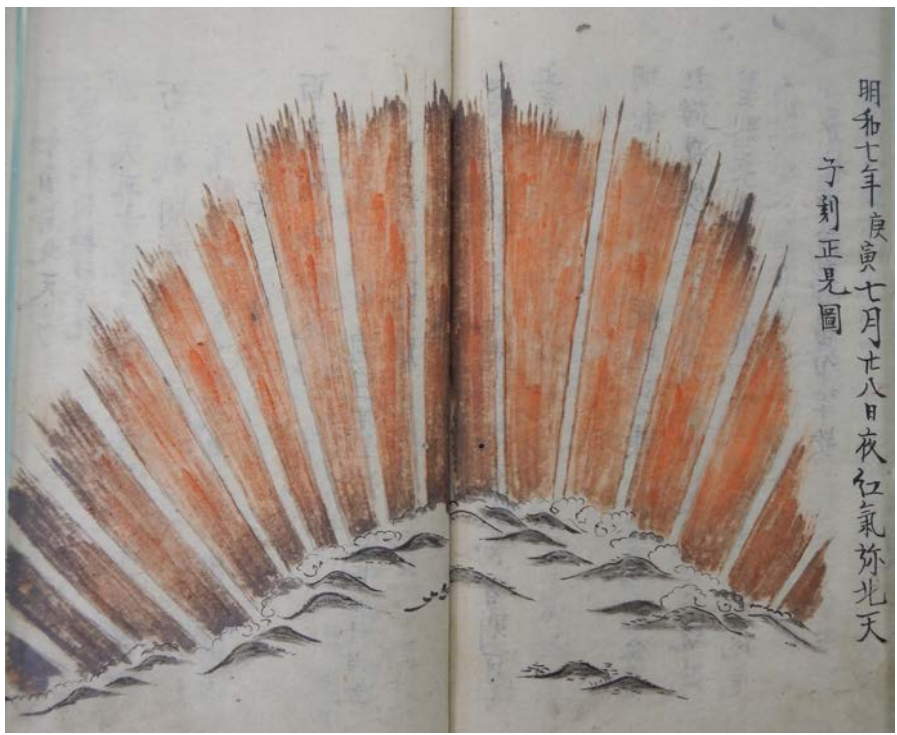

**図8.** 上から、日本書紀に記された620年の「赤氣」（国立公文書館所蔵；上）、『津和野百景図』第75図に見える雉尾（津和野町日本遺産センター所蔵：：中）、京都で観測された1770年の扇状のオーロラ（松阪郷土資料室所蔵：典籍64, ff. 25b-26a：下）、を各々示す。中段の図版に見える通り、飛翔中の雉の尾が開いて扇状になっているのに対し、他の雉の尾は閉じてアーク状になっている。

　774/775年の宇宙線イベント（Miyake *et al.* 2012）の前後ではメソポタミア北部のアミダで771/772年と773年6月にオーロラが見えたことが先述のズークニーン年代記に記録されている他（Hayakawa *et al.* 2017b）、776年にイングランドや中国で見られたオーロラを思わせる記述がアングロサクソン年代記や舊唐書・新唐書に記録されている（Usoskin *et al.* 2013）。このような歴史上のオーロラ様現象の記録にはしばしば大気光学現象が混入していることがあるので（Usoskin *et al.* 2017）、近代観測との慎重な比較検討が必要になる（Hayakawa *et al.* 2019b; Stephenson *et al.* 2019）。これらの記録は774/775年の宇宙線イベントの発生時期と必ずしも合致しないが（Büntgen *et al.* 2018）、少なくとも前後の活発な太陽活動を示唆していると考えられる（*e.g.*, Usoskin *et al.* 2013）。

　992/993 – 993/994年の宇宙線イベント（Miyake *et al.* 2012）周辺では特に992年末から993年初頭にかけてオーロラ記録が集中的に見受けられ、992年10月21日にザクセン、992年12月26日にザクセンとアルスター、992年12月〜翌1月に高麗でオーロラが記録されている。これらの記録と宇宙線イベントとの関係は依然熟考を要するが、774/775年周辺のオーロラ記録と同様、周囲での活発な太陽活動を裏付けるものと考えられる。

　太陽活動が活発な際は、2003年10月のハロウィーン・イベントや1921年5月の磁気嵐のようにCMEがしばしば連続して地球磁気圏に到達し（Newton 1948; Gopalswamy *et al.* 2005; Love *et al.*

2019b）、オーロラが数日続いて観測されることもある。本邦でも特に1204年2月21–23日の3晩に亘って「赤氣」が平安京周辺で観測されたことが『御室相承記』や『明月記』に記録されていることはしばしば議論されていた（神田 1933; 斉藤 1999）。同2月21日、中国側で肉眼黒点が『宋史』に記録されていたことから、Willis *et al.*（2005）はこの当時巨大黒点から複数のCMEが放出され、強烈な磁気嵐が3晩に亘って続いたことを指摘している。この現象は近年の研究で再度注目を浴びつつある（Kataoka *et al.* 2017）。

同様のことが1582年3月にも発生しており、3月6–8日に西欧一帯で、最終日の8日にはより磁気緯度の低い東アジア一帯でもオーロラが観測された（Hattori *et al.* 2019）。この宇宙天気現象時のオーロラオーバルの低緯度境界からDst値を推測すると−600 nTを下回るほどになり、1909年9月や1989年3月の極端磁気嵐に匹敵すると考えられる。

# 7 結論

以上、本節では歴史上の巨大宇宙天気現象について概観した。1859年9月のキャリントン・イベントは同時代の磁力観測やオーロラ記録、フレア、磁気嵐規模、オーロラの広がりなどの面で観測史上最大の宇宙天気現象と評価されていたが、1872年、1921年などの宇宙天気現象を近代観測から復元すると、これらも少なくとも磁気嵐規模やオーロラの広がりの面ではキャリントン・イベントと匹敵するものであったことが明らかになった。このような宇宙天気現象について国際地球観測年（1957年）以前のものは必ずしも定量的な評価が進んでいるわけではないので、今後さらなる復元研究が必要になる。

また、近代観測の時間幅を越えて宇宙天気現象の復元研究を行う場合、年輪や氷床コアの放射性同位体比に加えて、歴史文献中のオーロラ記録や黒点記録が特に過去のCMEについて手がかりをもたらすことがある。現状オーロラや黒点の記録は図像資料で過去9–12世紀、文字記録で過去22–27世紀近く遡るので、近代観測と歴史文献をつなぐことで、さらなる長期スパンでの極端宇宙天気現象の事例を追うことが可能となる。今年は奇しくも本邦で初めてオーロラが—いわば太陽地球圏の擾乱が—記録されたと考えられる620年より1400年の節目に当たる。この様な長期にわたって記録される過去の宇宙天気現象の定量的復元と現在の理論研究の照合は太陽地球系の長期変動を知るうえで重要であり、今後の発展が期待される。

# 参考文献

Akasofu, S.: The latitudinal shift of the auroral belt, *J. Atmos. Terr. Phys.*, **26**, 1167-1174, 1964.

Arlt, R.: Digitization of Sunspot Drawings by Staudacher in 1749-1796, *Solar Physics*, **247**, 399-410 (2008)

Baker, D. N., *et al*.: *Severe Space Weather Events— Understanding Societal and Economic Impacts*, Washington DC, National Academies Press, (2008).

Baker, D. N., *et al*.: A major solar eruptive event in July 2012: Defining extreme space weather scenarios, *Space Weather*, **11**, 585-591 (2013).

Barnard, L., McCracken, K. G., Owens, M. J., Lockwood, M.: What can the annual 10Be solar activity reconstructions tell us about historic space weather? *Journal of Space Weather and Space Climate*, **8**, A23 (2018).

Bartels, J.: Solar eruptions and their ionospheric effects—A classical observation and its new interpretation, *Terrestrial Magnetism and Atmospheric Electricity*, **42**, 235-239 (1937)

Beer, J., McCracken, K., von Steiger, R.: *Cosmogenic Radionuclides*, Berlin, Springer (2012).

Bicknell, P. J.: Did Anaxagoras observe a sunspot in 467 B.C.? *Isis*, **59**, 87 – 90 (1967).

Boteler, D. H.: The super storms of August/September 1859 and their effects on the telegraph system, Advances in Space Research, **38**, 159-172 (2006).

Büntgen, U., *et al*.: Tree rings reveal globally coherent signature of cosmogenic radiocarbon events in 774 and 993 CE, *Nature Communications*, **9**, 3605 (2018).

Carrasco, V. M. S., Aragonès, E., Ordaz, J., Vaquero, J. M.: The Great Aurora of January 1770 observed in Spain, History of Geo- and Space Sciences, **9**, 133-139 (2018).

Carrington, R. C.: Description of a Singular Appearance seen in the Sun on September 1, 1859, *Monthly Notices of the Royal Astronomical Society*, **20**, 13-15 (1859).

Chapman, S., Bartels, J.: *Geomagnetism, Vol. I: Geomagnetic and Related Phenomena*, London: Oxford Univ. Press (1940).

Chapman, S.: The Aurora in Middle and Low Latitudes, *Nature*, **179**, 7-11 (1957).

Clette, F., Svalgaard, L., Vaquero, J. M., Cliver, E. W.: Revisiting the Sunspot Number. A 400-Year Perspective on the Solar Cycle, *Space Science Reviews*, **186**, 35-103 (2014).

Cliver, E. W., Dietrich, W. F.: The 1859 space weather event revisited: limits of extreme activity, *Journal of Space Weather and Space Climate*, **3**, A31 (2013).

Cliver, E. W., Svalgaard, L.: The 1859 Solar-Terrestrial Disturbance And the Current Limits of Extreme Space Weather Activity, Solar Physics, **224**, 407-422 (2004).

Curto, J. J., Castell, J., Del Moral, F.: Sfe: waiting for the big one, Journal of Space Weather and Space Climate, **6**, A23 (2016).

Ebihara, Y., *et al*.: Statistical distribution of the storm‐time proton ring current: POLAR measurements, *Geophys. Res. Lett*., **29**, 1969 (2002).

Ebihara, Y., *et al*.: Possible Cause of Extremely Bright Aurora Witnessed in East Asia on 17 September 1770, *Space Weather*, **15**, 1373-1382 (2017).

Fujiyama, M., *et al*.: Revisiting Kunitomo's Sunspot Drawings During 1835 - 1836 in Japan, *Solar Physics*, **294**, 43 (2019).

Gonzalez, W. D., *et al*.: Interplanetary Origin of Intense, Superintense and Extreme Geomagnetic Storms, *Space Science Reviews*, **158**, 69-89 (2011).

Gopalswamy, N., *et al*.: Coronal mass ejections and other extreme characteristics of the 2003 October-November solar eruptions, *Journal of Geophysical Research: Space Physics*, **110**, A9, A09S15 (2005).

Green, J. L., Boardsen, S.: Duration and extent of the great auroral storm of 1859, *Advances in Space Research*, **38**, 130-135 (2006).

Hapgood, M. A.: *Space Weather: Its impact on Earth and implications for business*, Lloyd's, London (2010).

Hapgood, M. A.: The Great Storm of May 1921: An Exemplar of a Dangerous Space Weather Event, *Space Weather*, **17**, 950-975 (2019).

Hattori, K., Hayakawa, H., Ebihara, Y.: Occurrence of great magnetic storms on 6-8 March 1582, *Monthly Notices of the Royal Astronomical Society*, **487**, 3, 3550–3559 (2019).

Hayakawa, H., *et al*.: East Asian observations of low latitude aurora during the Carrington magnetic storm. *Publications of Astronomical Society of Japan*, **68**, 99 (2016b).

Hayakawa, H., *et al*.: Earliest datable records of aurora-like phenomena in the astronomical diaries from Babylonia. *Earth, Planets, and Space*, **68**, 195 (2016c).

Hayakawa, H., *et al*.: Historical aurora evidences for great magnetic storms in 990s. *Solar Physics*, **292**, 12 (2017a).

Hayakawa, H., *et al*.: The earliest drawings of datable auroras and a two-tail comet from the Syriac Chronicle of Zūqnīn. *Publications of Astronomical Society of Japan*, **69**, 17 (2017b).

Hayakawa, H., *et al*.: Records of Sunspots and Aurora Candidates in the Chinese Official Histories of the Yuán and Míng Dynasties during 1261-1644. *Publications of Astronomical Society of Japan*, **69**, 65 (2017c).

Hayakawa, H., *et al*.: Records of Auroral Candidates and Sunspots in Rikkokushi, Chronicles of Ancient Japan from Early 7th Century to 887, *Publications of Astronomical Society of Japan*, **69**, 86 (2017d).

Hayakawa, H., *et al*.: Long-lasting Extreme Magnetic Storm Activities in 1770 Found in Historical Documents, *The Astrophysical Journal Letters*, **850**, L31 (2017e).

Hayakawa, H., *et al*.: Iwahashi Zenbei's Sunspot Drawings in 1793 in Japan, *Solar Physics*, **293**, 8 (2018a).

Hayakawa, H., *et al*.: A great space weather event in February 1730, *Astronomy & Astrophysics*, **616**, A177 (2018c).

Hayakawa, H., *et al*.: The Great Space Weather Event during 1872 February Recorded in East Asia, *The Astrophysical Journal*, **862**, 15 (2018d).

Hayakawa, H., *et al*.: Low-Latitude Aurorae during the Extreme Space Weather Events in 1859, *The Astrophysical Journal*, **869**, 57 (2018e).

Hayakawa, H., *et al*.: The Extreme Space Weather Event in September 1909, *Monthly Notices of the Royal Astronomical Society*, **484**, 3, 4083-4099 (2019a).

Hayakawa, H., *et al*.: The Celestial Sign in the Anglo-Saxon Chronicle in the 770s: Insights on Contemporary Solar Activity. *Solar Physics*, **294**, 42 (2019b).

Hayakawa, H., *et al*.: A Comparison of Graphical Records in the East and the West, *Solar Physics*, **294**, 95 (2019c).

Hayakawa, H., *et al*.: Temporal and Spatial Evolutions of a Large Sunspot Group and Great Auroral Storms around the Carrington Event in 1859, *Space Weather* (2019e), DOI: 10.1029/2019SW002269.

Hayakawa, H., Mitsuma, Y., Ebihara, Y., Miyake, F.: The Earliest Candidates of Auroral Observations in Assyrian Astrological Reports: Insights on Solar Activity around 660 BCE, *The Astrophysical Journal Letters*, **884**, L18 (2019f).

Hodgson, R.: On a curious Appearance seen in the Sun, *Monthly Notices of the Royal Astronomical Society*, **20**, 15-16 (1859).

Hunger, H.: *Astrological reports to Assyrian kings*, Helsinki, Helsinki University Press (1992).

Kataoka, R., Iwahashi, K.: Inclined Zenith Aurora over Kyoto on 17 September 1770: Graphical Evidence of Extreme Magnetic Storm, *Space Weather*, **15**, 1314-1320 (2017).

Kataoka, R., *et al*.: Historical space weather monitoring of prolonged aurora activities in Japan and in China. *Space Weather*, **15**, 392-402 (2017).

Knipp, D. J., Fraser, B. J., Shea, M. A., Smart, D. F.: On the Little-Known Consequences of the 4 August 1972 Ultra-Fast Coronal Mass Ejecta: Facts, Commentary, and Call to Action, *Space Weather*, **16**, 1635-1643 (2018).

Korte, M., Constable, C.: Improving geomagnetic field reconstructions for 0-3 ka, *Physics of the Earth and Planetary Interiors*, **188**, 247-259 (2011).

Le, G., *et al*.: Morphology of the ring current derived from magnetic field observations, *Ann. Geophys.*, **22**, 1267–1295 (2004).

Lefèvre, L., *et al*.: Detailed Analysis of Solar Data Related to Historical Extreme Geomagnetic Storms: 1868 – 2010, *Solar Physics*, **291**, 1483-1531 (2016).

Lockyer, W. J. S.: Magnetic storm, 1909 Sept. 25, and associated solar disturbance, *Monthly Notices of the Royal Astronomical Society*, **70**, 12 (1909).

Love, J. J., Hayakawa, H., Cliver, E. W.: On the intensity of the magnetic superstorm of September 1909. *Space Weather*, 17, 37–45 (2019a).

Love, J. J., Hayakawa, H., Cliver, E. W.: Intensity and impact of the New York Railroad superstorm of May 1921, *Space Weather*, **17**, 1281–1292 (2019b).

McCracken, K. G., Dreschhoff, G. A. M., Zeller, E. J., Smart, D. F., Shea, M. A.: Solar cosmic ray events for the period 1561-1994: 1. Identification in polar ice, 1561-1950, *Journal of Geophysical Research*, **106**, A10, 21585-21598 (2001)

Mekhaldi, F., *et al.*: Multiradionuclide evidence for the solar origin of the cosmic-ray events of AD 774/5 and 993/4, Nature Communications, **6**, 8611 (2015).

Miyake, F., Nagaya, K., Masuda, K., Nakamura, T.: A signature of cosmic-ray increase in AD 774-775 from tree rings in Japan, *Nature*, **486**, 240-242 (2012).

Miyake, F., Masuda, K., Nakamura, T.: Another rapid event in the carbon-14 content of tree rings 2013, *Nature Communications*, **4**, 1748 (2013).

Newton, H. W.: "Sudden commencements" in the Greenwich magnetic records (1879–1944) and related sunspot data. *Geophysical Supplements to the Monthly Notices of the Royal Astronomical Society*, **5**, 159–185 (1948)

Ngwira *et al.*: Simulation of the 23 July 2012 extreme space weather event: What if this extremely rare CME was Earth directed? *Space Weather*, **11**, 671-679 (2013).

Owens, B.: Long-term research: Slow science, *Nature*, **495**, 300-303 (2013).

Riley, P., *et al.*: Extreme Space Weather Events: From Cradle to Grave, *Space Science Reviews*, **214**, 21 (2018).

Roach, F. E., Moore, J. G., Bruner, E. C., Jr., Cronin, H., Silverman, S. M.: The Height of Maximum Luminosity in an Auroral Arc, *Journal of Geophysical Research*, **65**, 3575 (1960).

Russell, C. T., Mewaldt, R. A., Luhmann, J. G., Mason, G. M., von Rosenvinge, T. T., Cohen, C. M. S., Leske, R. A., Gomez-Herrero, R., Klassen, A., Galvin, A. B., Simunac, K. D. C.: The Very Unusual Interplanetary Coronal Mass Ejection of 2012 July 23: A Blast Wave Mediated by Solar Energetic Particles, *The Astrophysical Journal*, **770**, 38 (2013).

Schrijver, C. J., *et al.*: Estimating the frequency of extremely energetic solar events, based on solar, stellar, lunar, and terrestrial records, *Journal of Geophysical Research: Space Physics*, **117**, A8, A08103 (2012).

Schröder, W.: The development of the aurora of 18 January 1770, *History of Geo- and Space Sciences*, **1**, 45-48 (2010).

Silverman, S. M.; Cliver, E. W.: Low-latitude auroras: the magnetic storm of 14-15 May 1921, *Journal of Atmospheric and Solar-Terrestrial Physics*, **63**, 523-535 (2001).

Silverman, S. M.: Low latitude auroras prior to 1200 C.E. and Ezekiel's vision, Advances in Space Research, **38**, 200-208 (2006).

Silverman, S. M.: Low-latitude auroras: The great aurora of 4 February 1872, *Journal of Atmospheric and Solar-Terrestrial Physics*, **70**, 1301-1308 (2008).

Siscoe, G. L., Silverman, S. M., Siebert, K. D.: Ezekiel and the Northern Lights: Biblical aurora seems plausible, Eos, Transactions American Geophysical Union, **83**, 173 (2002).

Smart, D. F., Shea, M. A., McCracken, K. G.: The Carrington event: Possible solar proton intensity time profile, *Advances in Space Research*, **38**, 215-225 (2006).

Stephenson, F. R., Willis, D. M., Hallinan, T. J.: Aurorae: The earliest datable observation of the aurora borealis, Astronomy & Geophysics, **45**, 6.15-6.17 (2004).

Stephenson, F. R., *et al*.: Do the Chinese Astronomical Records Dated A.D. 776 January 12/13 Describe an Auroral Display or a Lunar Halo? A Critical Re-examination, *Solar Physics*, **294**, 36 (2019).

Sugiura, M., Kamei, T.: Equatorial Dst index 1957‐1986, *IAGA Bull.*, **40**, ISGI Publication Office, Saint Maur des Fossess (1991).

Tsurutani, B. T., Gonzalez, W. D., Lakhina, G. S., Alex, S.: The extreme magnetic storm of 1-2 September 1859, *Journal of Geophysical Research Space Physics*, **108**, A7, 1268 (2003).

Usoskin, I. G., Kovaltsov, G. A.: Occurrence of Extreme Solar Particle Events: Assessment from Historical Proxy Data, *The Astrophysical Journal*, **757**, 92 (2012).

Usoskin, I. G., *et al*.: The AD775 cosmic event revisited: the Sun is to blame, *Astronomy & Astrophysics*, **552**, L3 (2013).

Usoskin, I. G., Kovaltsov, G. A., Mishina, L. N., Sokoloff, D. D., Vaquero, J.: An Optical Atmospheric Phenomenon Observed in 1670 over the City of Astrakhan Was Not a Mid-Latitude Aurora, *Solar Physics*, **292**, 15 (2017).

Usoskin, I. G.: A history of solar activity over millennia, *Living Reviews in Solar Physics*, **14**, 3 (2017).

Uusitalo, J., *et al*.: Solar superstorm of AD 774 recorded subannually by Arctic tree rings, *Nature Communications*, **9**, 3495 (2018).

Vaquero, J. M., Vázquez, M.: *The Sun Recorded Through History: Scientific Data Extracted from Historical Documents*, Berlin, Springer (2009).

Willis, D. M., Stephenson, F. R.: 2000

Willis, D. M., Stephenson, F. R.: Solar and auroral evidence for an intense recurrent geomagnetic storm during December in AD 1128, *Annales Geophysicae*, **19**, 289-302 (2001).

Willis, D. M., Davda, V. N., Stephenson, F. R.: Comparison between Oriental and Occidental Sunspot Observations, *Quarterly Journal of the Royal Astronomical Society*, **37**, 189 (1996a).

Willis, D. M., Stephenson, F. R., Singh, J. R.: Auroral Observations on AD 1770 September 16: the Earliest Known Conjugate Sightings, *Quarterly Journal of the Royal Astronomical Society*, **37**, 733 (1996b).

Willis, D. M., Armstrong, G. M., Ault, C. E., Stephenson, F. R.: Identification of possible intense historical geomagnetic storms using combined sunspot and auroral observations from East Asia, *Annales Geophysicae*, **23**, 945-971 (2005).

Willis, D. M., Stephenson, F. R., Fang, H.: Sporadic aurorae observed in East Asia, *Annales Geophysicae*, **25**, 417-436 (2007).

Willis, D. M., *et al*.: Sunspot observations on 10 and 11 February 1917: A case study in collating known and previously undocumented records. *Space Weather*, **16**, 1740-1752 (2018).

Wolff, E. W. *et al*.: The Carrington event not observed in most ice core nitrate records, *Geophysical Research Letters*, **39**, L08503 (2012).

Yau, K. K. C., Stephenson, F. R.: A revised catalogue of Far Eastern observations of sunspots (165 BC to AD 1918), *Royal Astronomical Society, Quarterly Journal*, **29**, 175-197 (1988).

Yokoyama, N., Kamide, Y., Miyaoka, H.: The size of the auroral belt during magnetic storms, *Annales Geophysicae*, **16**, 566-573 (1998).

神田茂:「本邦における極光の記録」,『天文月報』, **33**, 204-210 (1933).

斉藤国治: 『定家『明月記』の天文記録 : 古天文学による解釈』, 東京, 慶友社 (1999).

谷川清隆・相馬充: 「七世紀の日本天文学」, 『国立天文台報』, **11**, 31 - 55 (2008).

三津間康幸・早川尚志: 「世界最古のオーロラ文字記録と図像記録」, 『天文月報』, **110**, 472-479 (2017).

# Extreme Space Weather Events Recorded in History

Hisashi Hayakawa* and Yusuke Ebihara**

* Osaka University
** Kyoto University

## Abstract

This section shows an overview of a recent development of the studies on great space weather events in history. Its discussion starts from the Carrington event and compare its intensity with the extreme storms within the coverage of the regular magnetic measurements. Extending its analyses back beyond their onset, this section shows several case studies of extreme storms with sunspot records in the telescopic observations and candidate auroral records in historical records. Before the onset of telescopic observations, this section shows the chronological coverage of the records of unaided-eye sunspot and candidate aurorae and several case studies on their basis.